\documentclass[suppldata]{interact}
\usepackage{epstopdf}
\usepackage[caption=false]{subfig}

\usepackage[numbers,sort&compress,merge]{natbib}
\bibpunct[, ]{[}{]}{,}{n}{,}{,}
\renewcommand\bibfont{\fontsize{10}{12}\selectfont}

\usepackage{amsmath,amsfonts,amsthm}
\usepackage[dvipsnames]{xcolor}
\usepackage{graphicx}
\usepackage{multirow,chemformula}
\usepackage[version=4]{mhchem}
\usepackage[colorlinks=true,breaklinks=true,linkcolor=magenta,citecolor=magenta,urlcolor=magenta]{hyperref}
\usepackage{comment}
\usepackage{array}
\usepackage{booktabs}
\usepackage{tablefootnote}

\theoremstyle{plain}

\theoremstyle{definition}

\theoremstyle{remark}

\newcommand{\psitrial}{\psi_\mathrm{T}}

\newcommand{\op}[1]{\hat{#1}}
\newcommand{\crt}[1]{\op{a}^\dagger_{#1}}
\newcommand{\dst}[1]{\op{a}^{\phantom{\dagger}}_{#1}}
\newcommand{\hamiltonian}[1]{\op{H}_{#1}}
\newcommand{\hinitial}{\hamiltonian{\mathrm{i}}}
\newcommand{\hfinal}{\hamiltonian{\mathrm{f}}}
\newcommand{\aspham}[1]{\op{H}(#1)}
\newcommand{\parent}{\op{H}_{\mathrm{P}}}
\newcommand{\protoparent}{\parent^*}

\newcommand{\pauliset}{\mathcal{S}}
\newcommand{\pauli}[1]{\op{P}_{#1}}
\newcommand{\kerspace}{\mathrm{Ker}}

\newcommand{\variance}[2]{\mathrm{Var}_{#1}[#2]}
\newcommand{\vett}[1]{\boldsymbol{#1}}
\newcommand{\alphavec}{\vett{\alpha}}
\newcommand{\betavec}{\vett{\beta}}
\newcommand{\spectro}[2]{${}^{#1}{#2}$}
\newcommand{\aolabel}[1]{$#1$}
\newcommand{\TIME}[1]{T_{\mathrm{#1}}}

\newcommand{\revision}[1]{{#1}}

\begin{document}

\title{Adiabatic state preparation from general initial states}

\author{
\name{
Bryce Fuller\textsuperscript{a*}\thanks{\textsuperscript{*}bryce.fuller@ibm.com},
Mario Motta\textsuperscript{a},
Stuart M. Harwood\textsuperscript{b},
Chetan Murthy\textsuperscript{c},
Tanvi P. Gujarati\textsuperscript{c},
Antonio Mezzacapo\textsuperscript{a},
Dimitar Trenev\textsuperscript{b**}\thanks{\textsuperscript{**}dimitar.trenev@exxonmobil.com}}
\affil{
\textsuperscript{a} IBM Quantum, IBM T. J. Watson Research Center, Yorktown Heights, NY 10598; \\
\textsuperscript{b} ExxonMobil Corporate Strategic Research, Annandale, NJ 08801, USA; \\
\textsuperscript{c} IBM Quantum, IBM Research Almaden, San Jose, CA 95120}
}

\maketitle

\begin{abstract}
A variety of quantum computing algorithms exist for the preparation of approximate Hamiltonian ground states. A natural and important question is how these ground-state approximations can be further improved using adiabatic state preparation.
Here, we present a heuristic method to carry out adiabatic state preparation starting from a generic initial wavefunction. Given a quantum circuit that prepares the initial wavefunction, and a target Hamiltonian for which one wishes to prepare the ground state, we present an algorithm to construct an adiabatic path between these two states. This method works by approximating a parent Hamiltonian for the initial wavefunction, and the quality of this approximation can be can be checked prior to running the ASP algorithm. 
We apply this technique to simulate the ground state of water and the lowest-lying multireference singlet state of methylene, using various initial wavefunctions as the starting point of the adiabatic path.
\end{abstract}

\begin{keywords}
Quantum computation; parent Hamiltonian, adiabatic state preparation
\end{keywords}

\section{Introduction}

Solving the many-particle Schr\"odinger equation to compute eigenpairs of a Hamiltonian is an important application for a quantum computer. 
In particular, it arises in the simulation of the electronic structure of molecules and materials \cite{cao2019quantum,mcardle2020quantum,bauer2020quantum,cerezo2021variational}.

A prominent quantum computing algorithm for preparing Hamiltonian ground states is adiabatic state preparation (ASP) \cite{farhi2000quantum,farhi2001quantum,aspuru2005simulated,babbush2014adiabatic, veis2014adiabatic,albash2018adiabatic}. Within ASP, one slowly varies the Hamiltonian of a quantum 
computer starting with a simple operator  $\hinitial$, whose ground state is known, and ending with a final operator $\hfinal$, whose ground state one wishes to prepare.
If the adiabatic evolution starts from the ground state of $\hinitial$, the Hamiltonian is transformed sufficiently slowly, and the energy gap between the ground and the lowest-excited state remains nonzero across the adiabatic evolution, then the register 
remains in the ground state of the instantaneous Hamiltonian according to the adiabatic theorem \cite{born1928beweis,kato1950adiabatic,messiah1999quantum,avron1998adiabatic,teufel2003adiabatic}.
ASP can thus be used as a heuristic method for ground-state approximation in quantum chemistry and, more generally, in many-body quantum mechanics.
The adiabatic evolution often starts from a relatively crude ground-state approximation.
In electronic structure, this approximation is often a Hartree–Fock (HF) or a Density Functional Theory (DFT) wavefunction. For these wavefunctions, the ``parent'' Hamiltonian $\hinitial$ is a simple one-body operator. \revision{ASP is known to be unsuitable when applied to highly correlated systems such as those with strong multireference characteristics, however, there have been promising results to indicate that a careful choice of initial state, scheduling function, and penalty term can improve the stability of ASP in the strongly correlated regime \cite{asp2022correlated}.}

In recent years, a wealth of heuristic quantum algorithms producing ground-state approximations have been designed and demonstrated on classical simulators of quantum hardware \cite{kandala2017hardware,google2020hartree,huggins2022unbiasing,zhao2023orbital,motta2023quantum,weaving2023contextualVQE_N2}.
These quantum algorithms produce trial wavefunctions $\psitrial$ that, in principle, can be used as a starting point for ASP. To achieve this goal, however, it is necessary to produce a parent Hamiltonian for $\psitrial$, i.e. a Hermitian operator $\hinitial$ whose ground state is $\psitrial$ \cite{fannes1992finitely}.

In this work, we introduce a heuristic algorithm to construct an adiabatic path starting at any wavefunction that can be prepared efficiently on a quantum computer. The algorithm begins with a series of measurements to the starting state, and uses this information to build a parent Hamiltonian for the initial wavefunction. Our method sometimes identifies a continuous subspace of valid parent Hamiltonians. In these cases, one has the opportunity to impose additional constraints onto the parent Hamiltonian's structure. This can be used to optimize for practical considerations such as sparsity or similarity metrics between the parent and target Hamiltonians.  

The remainder of this work is structured as follows: in Section \ref{subsec:asp}, we briefly review the ASP method and present the algorithm to construct a parent Hamiltonian for a generic trial wavefunction. In Section \ref{sec:result}, we apply the proposed algorithm to simulate the ground state of water and the lowest-lying multireference singlet state of methylene. Results are discussed in Section~\ref{sec:discussion}, and conclusions are drawn in Section \ref{sec:conclusion}.

\section{Methods}
\label{sec:method}

We now describe the methods used in this work. After a brief review of adiabatic state preparation (ASP) in \ref{subsec:asp} we describe how the construction of a parent Hamiltonian is converted into a combinatorial optimization problem which can then be approximated using numerical optimization methods. 

\subsection{Adiabatic state preparation}\label{subsec:asp}

Within ASP \cite{farhi2000quantum,farhi2001quantum,aspuru2005simulated,babbush2014adiabatic, veis2014adiabatic,albash2018adiabatic}, a quantum state of $n$ qubits evolves in time according to 
the Schr\"{o}dinger equation
\begin{equation}
i \frac{d}{dt} | \Psi(t) \rangle = \aspham{t} | \Psi(t) \rangle 
\;,\;
0 \leq t \leq \TIME{ASP} 
\;,
\end{equation}
or equivalently
\begin{equation}
\label{eq:asp}
\frac{i}{\TIME{ASP}} \frac{d}{ds} | \Psi(s) \rangle = \aspham{s} | \Psi(s) \rangle 
\;,\;
0 \leq s \leq 1
\;.
\end{equation}
In Eq.~\eqref{eq:asp}, $\aspham{s}$ is a time-dependent Hamiltonian operator 
that interpolates between $\aspham{0} = \hinitial$ and
$\aspham{1} = \hfinal$,
and $|\Psi(0) \rangle = | \psitrial \rangle$
is the ground state of $\hinitial$.
One of the possible choices for the time-dependent Hamiltonian
is the linear interpolation
\begin{equation}
\label{eq:linear_interpol}
\aspham{s} = (1-s) \hinitial + s \hfinal
\;.
\end{equation}
According to the adiabatic theorem \cite{born1928beweis,kato1950adiabatic,messiah1999quantum,avron1998adiabatic,teufel2003adiabatic}, when the time-dependent Hamiltonian is gapped and the evolution time $\TIME{ASP}$ is sufficiently high, the evolution Eq.~\eqref{eq:asp} follows the instantaneous ground state of $\aspham{s}$, and therefore $| \Psi(1) \rangle$ accurately approximates the ground state of $\aspham{1} = \hfinal$. More precisely, denoting
$\aspham{s} = \sum_j \lambda_j(s) | \phi_j(s) \rangle \langle \phi_j(s) |$ the spectral decomposition of $\aspham{s}$ with eigenvalues in ascending order, one has 
the following bound \cite{jordan2008quantum} for the distance between the final state of the ASP, $| \Psi(s=1) \rangle$, and the ground state of the target Hamiltonian,
$| \phi_0(s=1) \rangle$,

\begin{equation}
\Big\| | \Psi(1) \rangle - | \phi_0(1) \rangle \Big\| \leq \frac{1}{\TIME{ASP}}
\left[
\frac{f_{1}(1)}{\gamma^{2}(1)} +
\frac{f_{1}(0)}{\gamma^{2}(0)} +
\int_{0}^{1}
\left[
\frac{5f_{1}^{2}(s)}{\gamma^{3}(s)} + 
\frac{f_{2}(s)}{\gamma^{2}(s)} 
\right] ds 
\right]
\;,
\end{equation}
where 
\begin{equation}
f_{k}(s) = \left\| \frac{d^k \hat{H}}{ds^k}(s) \right\|
\;,\;
\gamma(s) = \lambda_1(s) - \lambda_0(s)
\;.
\end{equation}
A quantitative measure of the time required for a stable evolution is given by the following adiabatic estimate \cite{lee2023evaluating, amin2009consistency}
\begin{equation}
\label{eq:t_est}
\TIME{est} = \max_{0 \leq s \leq 1} \max_{j>0}
\frac{
\left| \langle \phi_j(s) | \frac{d \hat{H}}{ds}(s) | \phi_0(s) \rangle \right|}
{ \left( \lambda_j(s) - \lambda_0(s) \right)^2 }
\;,
\end{equation}
that we will use throughout our work \revision{to benchmark adiabatic paths resulting from different choices of parent Hamiltonian. This quantity can be numerically estimated by diagonalizing $\aspham{s}$ at a discretized set of $s$ values. We note that $\TIME{est}$ does not depend explicitly on the gradient of s and is therefore independent of one's choice of scheduling function. 
}

\subsection{Construction of a parent Hamiltonian}
\label{sec:parent_and_asp}

The main contribution of this work is a method to approximately construct
a parent Hamiltonian for a broad class of trial wavefunctions $\psitrial$.
Given access to a quantum circuit that prepares $\psitrial$, 
this section details how to approximately construct a parent Hamiltonian for this wavefunction, i.e. a Hermitian operator $\parent$ such that $\psitrial$ lies in the ground eigenspace of $\parent$.
This is done by (i) constructing a family of proto-parent\footnote{A proto-parent Hamiltonian for $\psitrial$ is a Hermitian operator having $\psitrial$ as an eigenstate, though not necessarily the ground state} Hamiltonians $\protoparent[\alphavec]$ for $\psitrial$, parameterized by a vector $\alphavec \in \mathbb{R}^m$, (ii) applying a transformation that folds the spectrum of $\protoparent[\alphavec]$ to obtain a parent Hamiltonian $\parent[\alphavec]$ for any choice of $\alphavec$, and finally
(iii) optimizing the free parameters $\alphavec$ to ensure $\parent[\alphavec]$ is as close as possible to $\hfinal$.

\subsubsection{From a proto-parent to a parent Hamiltonian}

We start from a set $\pauliset$ of $m$ non-identity Pauli operators acting on $n$ qubits. We will search for a proto-parent Hamiltonian of the form $\protoparent[\alphavec] = \sum_{i=1}^m \alpha_i \pauli{i}$. In this work, we map the target system's second-quantized Born-Oppenheimer electronic Hamiltonian into a linear combination of $n$-qubit Pauli operators, $\hfinal = \sum_{i=1}^{m} \beta_i \pauli{i}$, and use the resulting set of non-identity Pauli operators to define $\pauliset = \{ \pauli{i} \}_{i=1}^m$.
To find the vector $\alphavec \in \mathbb{R}^m$, we use the zero-variance property of eigenstates, i.e.
\begin{equation}
0 = \variance{{\psitrial}}{\protoparent[\alphavec]} = \langle \psitrial | \protoparent[\alphavec]^2 | \psitrial \rangle - \langle \psitrial | \protoparent[\alphavec] | \psitrial \rangle^2
\;.
\end{equation}
Using the equation $\protoparent[\alphavec] = \sum_i \alpha_i \pauli{i}$, we readily obtain

\begin{equation}
\begin{split}
\label{eq:a_matrix}
0 = \sum_{ij} \alpha_i \alpha_j 
&\Big[ 
\langle \psitrial | \pauli{i} \pauli{j} | \psitrial \rangle 
- 
\langle \psitrial | \pauli{i} | \psitrial \rangle
\langle \psitrial | \pauli{j} | \psitrial \rangle
\Big] \\
= \sum_{ij} \alpha_i \alpha_j 
&\left[ 
\langle \psitrial | \frac{\{ \pauli{i} , \pauli{j} \}}{2}  | \psitrial \rangle 
- 
\langle \psitrial | \pauli{i} | \psitrial \rangle
\langle \psitrial | \pauli{j} | \psitrial \rangle
\right] \\
\equiv
\sum_{ij} \alpha_i \alpha_j &A_{ij}
= \alphavec \cdot (A \alphavec)
\end{split}
\end{equation}

In the second equality, we replaced the complex-valued quantity $\langle \psitrial | \pauli{i} \pauli{j} | \psitrial \rangle$ with its real part $\langle \psitrial | \{ \pauli{i} , \pauli{j} \} | \psitrial \rangle/2$. 
Since $A$ is a covariance matrix between Pauli operators it is non-negative\footnote{for a generic vector $\vett{x}$, one has $\vett{x} \cdot (A \vett{x}) = \variance{{\psitrial}}{\protoparent[\vett{x}]} \geq 0$}, and due to the non-negativity property, $\protoparent[\alphavec]$ is a proto-parent Hamiltonian for $\psitrial$ if and only if $\alphavec \in \kerspace(A)$. 
If $A$ has a non-trivial kernel, a subspace of proto-parent Hamiltonians has been found. Otherwise, the procedure is unsuccessful. This situation can be identified very easily, and can be handled in several ways, e.g. 
(i) extending the set $\pauliset$ and evaluating additional rows/columns in a larger $A$ matrix until a nontrivial kernel is found, (ii) retaining eigenvectors of $\kerspace(A)$ with eigenvalues below an user-defined threshold, or (iii) considering a different set $\pauliset$ constructed not from the Hamiltonian but, e.g., from Pauli generators in the quantum circuit preparing $\psitrial$.

While $\psitrial$ is an eigenstate of $\protoparent[\alphavec]$ with eigenvalue $\lambda[\alphavec] = \langle \psitrial | \protoparent[\alphavec] | \psitrial \rangle$, it is not necessarily in the ground eigenspace of this operator.
To obtain a parent Hamiltonian for $\psitrial$, we employ a spectrum folding operation, commonly used in numerical simulations of specific excited states \cite{umrigar1988optimized,hanscam2022applying,otis2023promising,tazi2024folded},
\begin{equation}
\label{eq:parent}
\parent[\alphavec] 
= 
\left(  \protoparent[\alphavec] - \lambda[\alphavec] \op{I}
\right)^2
=
\sum_{ij} \alpha_i \alpha_j \pauli{i} \pauli{j}
+ ( \alphavec \cdot \vett{b} )^2 \op{I}
- 2 ( \alphavec \cdot \vett{b} ) \sum_i \alpha_i \pauli{i}
\;,
\end{equation}
where $b_i = \langle \psitrial | \pauli{i} | \psitrial \rangle$. Note that we used $\alphavec \cdot \vett{b} = \lambda[\alphavec]$ in Eq.~\eqref{eq:parent}.

\subsubsection{Optimization of the $\alphavec$ vector}\label{alpha_optimization}

While Eq.~\eqref{eq:parent} yields a parent Hamiltonian for $\psitrial$, the vector $\alphavec$ is still to be determined. 
To this end, we propose a heuristic protocol: we seek for an $\parent[\alphavec]$ as close as possible to $\hfinal$ by minimizing the following cost function,
\begin{equation}
\label{eq:cost}
C[\alphavec] = \| \parent[\alphavec] - \hfinal \|_\mathrm{F}^2
\;,\;
\alphavec \in \kerspace(A)
\;,
\end{equation}
where $\| \hat{X} \|_\mathrm{F}^2 = \mbox{Tr}[ \hat{X}^\dagger \hat{X} ]$ is the Frobenius norm.
The cost function is quartic in $\alphavec$, which makes it expensive to evaluate and challenging to optimize. To circumvent such difficulties, we approximate the cost function by: (i) penalizing Pauli operators that lead to non-identity terms in $\parent^*[\alphavec]^2$
and (ii) removing terms contributing only an energy shift. The resulting approximation justified in Appendix \ref{app:cost_function}, is:

\begin{equation}
\label{eq:cost_simplified}
C[\alphavec] \mapsto \| 2 (\vett{b} \cdot \alphavec) \alphavec + \betavec \|^2 + \rho \sum_{ij} \alpha_i \alpha_j
\bar\delta_{ij}
\;,\;
\alphavec \in \kerspace(A)
\;,
\end{equation}
where $\betavec \in \mathbb{R}^m$ is the vector of coefficients that defines $\hfinal$, $\rho \geq 0$ is a penalty hyper-parameter, and $\bar\delta_{ij}=1$ if and only if $i \neq j$ and $[ \pauli{i}, \pauli{j} ]=0$. Notwithstanding the simplification of the cost function, its minimization remains a non-convex optimization problem, non-trivial to initialize and converge. 
We start the optimization from a vector $\alphavec^*$ computed by minimizing $\| 2 (\vett{b} \cdot \alphavec) \alphavec + \betavec \|^2$ under the additional constraint that $\vett{b} \cdot \alphavec$ is fixed. This permits an explicit solution of the resulting least-squares problem. The formulation of $\alphavec^*$ us discussed further in Appendix \ref{app:cost_initial}.

\subsection{Trial wavefunctions}

In this work, we employ various trial wavefunctions as starting points for ASP. Of the states listed below, only the RHF has a parent Hamiltonian that is a simple one-body operator (the Fock operator, see also Section \ref{sec:result}).
\begin{enumerate}
\item The restricted closed-shell Hartree-Fock (RHF) state. Throughout this work, we use RHF orbitals as the one-electron basis set. With this choice, using standard qubit mappings (e.g. Jordan-Wigner or Bravyi-Kitaev, with or without qubit tapering \cite{seeley2012bravyi,bravyi2017tapering}), the RHF state is encoded with a single computational basis state, $| \psi_{\mathrm{RHF}} \rangle = | \vett{x}_{\mathrm{RHF}} \rangle$.
\item The Moller-Plesset second-order perturbation theory (MP2) state,
\begin{equation}
| \psi_{\mathrm{MP2}} \rangle = \nu 
\left[ 1 + \sum_{aibj,\sigma\tau} t^{(2)}_{aibj,\sigma\tau} \crt{a\sigma}\crt{b\tau}\dst{j\tau}\dst{i\sigma}
\right]
| \psi_{\mathrm{RHF}} \rangle
\;,
\end{equation}
where $ij$/$ab$ label occupied/virtual orbitals in the RHF state, $\dst{i\sigma}$/$\crt{a\tau}$ is the operator destroying/creating an electron with spin $\sigma$/$\tau$ from orbital $i$/$a$, the $t_2$ amplitudes are obtained on a classical computer, and $\nu$ is a normalization factor. The MP2 state is a linear combination of $O(N_e^2N_o^2)$ computational basis states, where $N_e$/$N_o$ is the number of electrons/orbitals in the molecule under study. Such a wavefunction can be prepared with, e.g. the algorithms proposed by Tubman et al \cite{tubman2018postponing} or Parrish et al \cite{parrish2019quantumMC}.
\item A Ritz vector from a $d$-dimensional Hamiltonian Krylov subspace,
\begin{equation}
\label{eq:krylov}
| \psi^{\mathrm{Krylov}}_{d} \rangle = \sum_{p=0}^{d-1} c_p \hat{H}^p | \psi_{\mathrm{RHF}} \rangle
\;,
\end{equation}
i.e., the lowest-energy state in the $d$-dimensional Krylov space spanned by the vectors $\hat{H}^p | \psi_{\mathrm{RHF}} \rangle$ with $p=0 \dots d-1$. The Ritz vector Eq.~\eqref{eq:krylov} can be computed e.g. with the algorithms proposed by Kirby et al~\cite{kirby2023exact} and Seki et al \cite{seki2021quantum}, focused on the Hamiltonian Krylov space. An alternative to  Eq.~\eqref{eq:krylov} is to employ powers of the time-evolution operator $\exp(-i \Delta t \hat{H})$ rather than of $\hat{H}$ to generate a Krylov space \cite{parrish2019quantum,cohn2021quantum,klymko2022real,motta2023subspace}.
\end{enumerate}

\subsection{Computational details}
\label{sec:details}

The overall strategy for the calculations performed in this work involved initial pre-processing of the molecule descriptions to generate optimized RHF orbitals, active-space Hamiltonians, and trial wavefunctions. After the pre-processing phase, we compute the parent Hamiltonian using the strategy given in previous subsections. For the numerics shown in this work, the expectation values of all elements of $\pauliset$ were computed with classical noiseless simulations and the adiabatic estimate in Eq.~\eqref{eq:t_est} is calculated via classical diagonalization of $\aspham{s}$. Additional details for the studied molecules are listed in Appendix~\ref{eq:app_chem}.

We performed closed-shell RHF and MP2 calculations at STO-6G level of theory using PySCF \cite{sun2018pyscf,sun2020recent}.
We mapped the active-space Hamiltonian onto a qubit operator using the Jordan-Wigner transformation, as implemented in IBM's open-source Python library for quantum
computing, Qiskit \cite{qiskit2023}. Qiskit provides tools for various tasks such as creating and manipulating operators 
expressed as linear combinations of Pauli operators.
To leverage the conservation of particle number modulo 2 and other $\mathbb{Z}_2$ symmetries of molecular 
orbitals, we employed qubit tapering \cite{bravyi2017tapering,mishmash2023hierarchical}. Additional information about the treatment of symmetries in the search for parent Hamiltonians is given in the Appendix.

\section{Results}
\label{sec:result}

\begin{figure}
\centering
\includegraphics[width=0.8\textwidth]{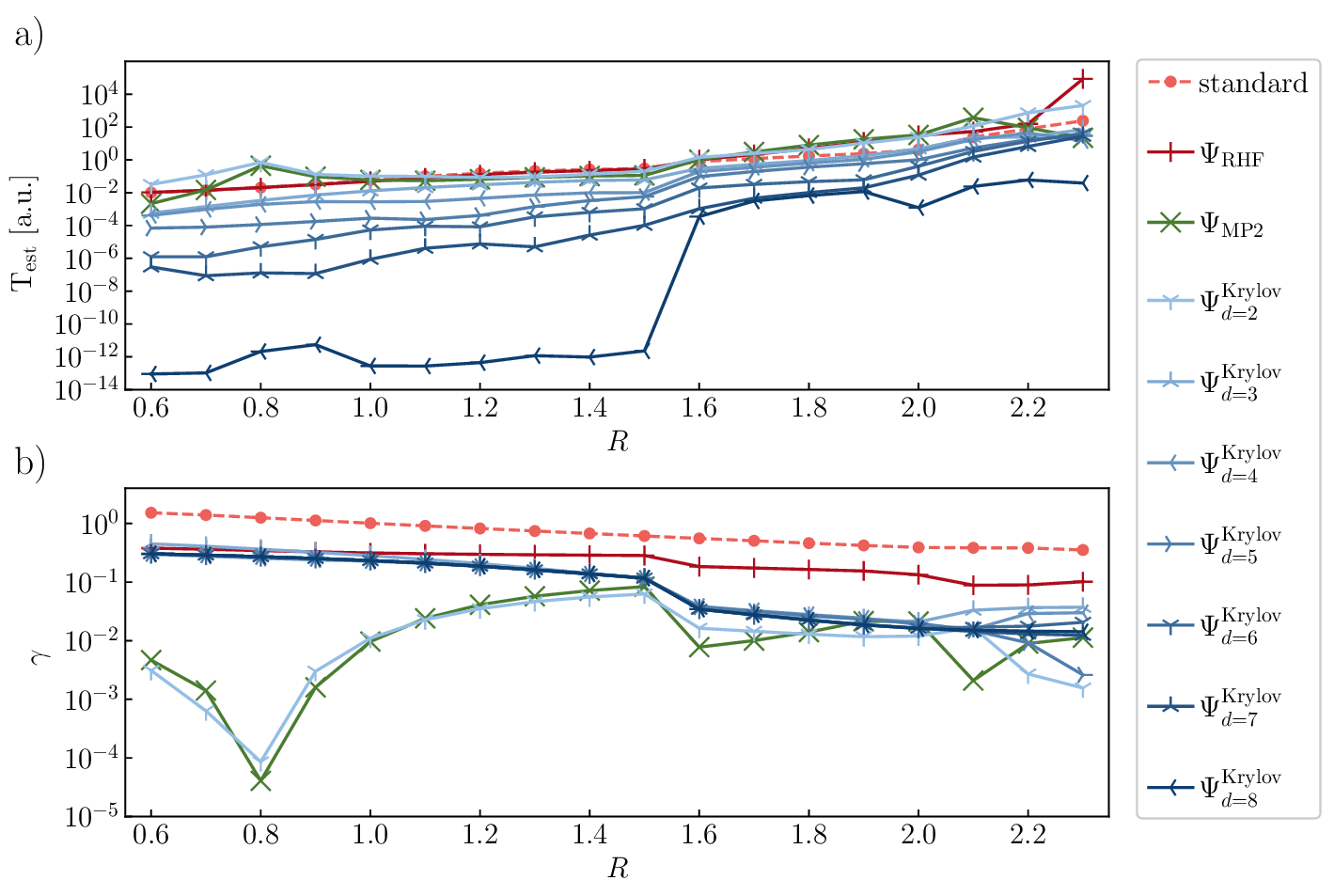}
\caption{ 
\textbf{(a)} Estimate of the adiabatic time $\TIME{est}$ along dissociation of both OH bonds in the H$_2$O molecule, using a (4e,4o) active space. $\TIME{est}$ is estimated for standard adiabatic state preparation (i.e. with RHF trial and the Fock operator as its parent, red circles connected by dashed lines) and for different trial wavefunctions (red,  green, blue symbols for RHF, MP2, Krylov-space wavefunctions respectively) in combination with the parent Hamiltonian construction proposed in this work. \textbf{(b)} Spectral gap $\gamma$ of the parent Hamiltonians obtained by applying our method to various trial wavefunctions, compared against the spectral gap of the Fock operator used in standard adiabatic state preparation. For both subplots, $R$ is a dimensionless quantity which scales the length of all molecular bond lengths relative to the equillibrium geometry. 
}
\label{fig:h2o_4e4o}
\end{figure}

In this Section, we present results for the dissociation of water and for singlet methylene. We compare the proposed  protocol with ``standard ASP''. By this term, we mean ASP using the RHF state as trial wavefunction and the Fock operator,
\begin{equation}
\hat{F} = \sum_{p\sigma} \varepsilon_p \, \crt{p\sigma} \dst{p\sigma}
\;
\end{equation}
where $\varepsilon_p$ are the molecular orbital (MO) energies,
as the parent Hamiltonian \cite{veis2014adiabatic}.

\subsection{Dissociation of water}

\begin{figure}
\centering
\includegraphics[width=0.8\textwidth]{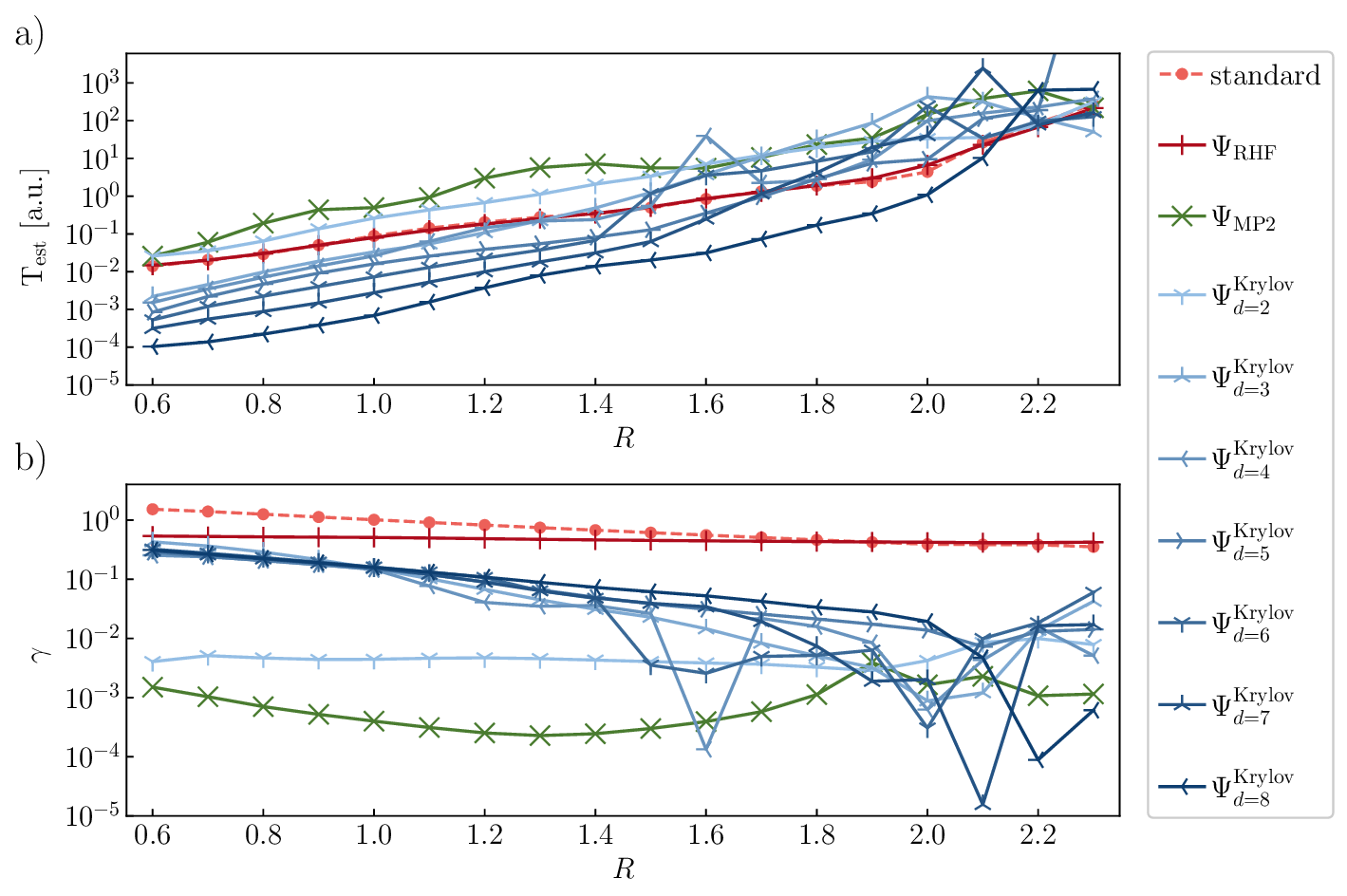}
\caption{
\textbf{(a)} Estimate of the adiabatic time $\TIME{est}$ along dissociation of both OH bonds in the H$_2$O molecule, using a (8e,6o) active space. $\TIME{est}$ is estimated for standard adiabatic state preparation (i.e. with RHF trial and the Fock operator as its parent, red circles connected by dashed lines) and for different trial wavefunctions (red,  green, blue symbols for RHF, MP2, Krylov-space wavefunctions respectively) in combination with the parent Hamiltonian construction proposed in this work. \textbf{(b)} Spectral gap $\gamma$ of the parent Hamiltonians obtained by applying our method to various trial wavefunctions, compared against the spectral gap of the Fock operator used in standard adiabatic state preparation. For both subplots, $R$ is a dimensionless quantity which scales the length of all molecular bond lengths relative to the equillibrium geometry
}
\label{fig:h2o_8e6o}
\end{figure}

We consider the dissociation of both OH bonds of the H$_2$O molecule, in a (4e,4o) active space of MOs with \aolabel{2p_x} and \aolabel{2p_y} (in-plane) character for oxygen and \aolabel{1s} character for hydrogen, and a (8e,6o) active space of MOs with \aolabel{2s}, \aolabel{2p} character for oxygen and \aolabel{1s} character for hydrogen.
We calculated the adiabatic estimate $\TIME{est}$ over a range of scaled bond distances, with 5 and 8 qubits respectively. These results are illustrated in Figures \ref{fig:h2o_4e4o} and \ref{fig:h2o_8e6o} for the (4e,4o) and (8e,6o) active spaces respectively.

The increase in $\TIME{est}$ between 0.6 and 2.3 $\mathrm{\AA}$ is consistent with the fact that electron correlation (and correspondingly qubit entanglement) strengthens as the OH bond lengths increase.
The adiabatic estimate is seen to increase monotonically with bondlength for ``standard'' ASP. A similar trend is seen in Figure~\ref{fig:h2o_4e4o}b for ASP with various trial wavefunctions and the parent Hamiltonian proposed in this work, albeit with fluctuations.
We observe that the adiabatic estimates for paths starting in a Krylov initial state decrease with Krylov space dimension $d$ (light to dark blue), despite the fact that the parent Hamiltonians have a decreasing spectral gap. This is consistent with observing that increased $d$ yields trial wavefunctions that have higher overlap with the exact ground state.

Similar behavior is seen in Figure \ref{fig:h2o_8e6o}, where $\TIME{est}$ for ASP with RHF trial state and parent Hamiltonian minimizing Eq.~\eqref{eq:cost} essentially tracks the ``standard'' counterpart. For Krylov initial states, $\TIME{est}$ decreases with Krylov space dimension $d$ for bondlengths $R\leq1.4$ $\mathrm{\AA}$. For $R>1.4$ we observe noticeable fluctuations coinciding with sharp decreases in the spectral gap of the parent Hamiltonian. 

\revision{As $R$ increases away from its equilibrium value, the ground state of H$_2$O becomes more highly correlated and less suited to ASP. It is a general phenomena that as systems become increasingly multireference in character, the cost of ground state preparation will begin to approach the worst case instances which earn this problem its QMA-Hard complexity. Encouragingly, there have been efforts to optimize the performance of ASP for strongly correlated systems by using particular initial states and including penalty terms into the time dependent Hamiltonian \cite{asp2022correlated}. }

\subsection{Singlet methylene}

We now consider the lowest-energy \spectro{1}{\mathrm{A}_1} state of methylene, CH$_2$.
While the ground state of methylene is a singlet-reference \spectro{3}{\mathrm{B}_1} state, 
the \spectro{1}{\mathrm{A}_1} state is well-known for its multireference character,
which is a consequence of the near-degeneracy of the boundary \spectro{3}{a_1} and \spectro{1}{b_1} MOs.
The \spectro{1}{\mathrm{A}_1} state, in the (2e,2o) active space spanned by the boundary MOs, is 
\begin{equation}
| \Psi_{\textrm{\spectro{1}{\mathrm{A}_1}}} \rangle = \left[ \cos(\varphi) \crt{h\uparrow} \crt{h\downarrow} - \sin(\varphi) \crt{l\uparrow} \crt{l\downarrow} \right] | \emptyset \rangle
\;,\; \varphi \simeq 0.20
\;,\;
\end{equation}
where $|\emptyset \rangle$ is the vacuum state and $h$/$l$ labels the \spectro{3}{a_1}/\spectro{1}{b_1} MO,
whereas the triplet state is 
\begin{equation}
| \Psi_{\textrm{\spectro{3}{\mathrm{B}_1}}} \rangle = \crt{l\uparrow} \crt{h\uparrow} | \emptyset \rangle
\;.
\end{equation}
Elucidating the electronic structure of methylene was one of the early successed of theoretical chemistry \cite{harrison1971electronic,o1971c,hay1972generalized,schaefer1986methylene,goddard1985theoretical,shavitt1985geometry,lineberger2011synergy}, 
and its \spectro{1}{\mathrm{A}_1} state remains a compelling system to test computational methods \cite{pittner1999assessment,evangelista2006high,bhaskaran2010multireference,demel2008multireference}.
Since the \spectro{1}{\mathrm{A}_1} state can be prepared by ASP starting with a singlet initial state, 
e.g. the RHF state, it is a relevant test case for the parent Hamiltonians introduced in the present work.

\begin{figure}
\centering
\includegraphics[width=0.6\textwidth]{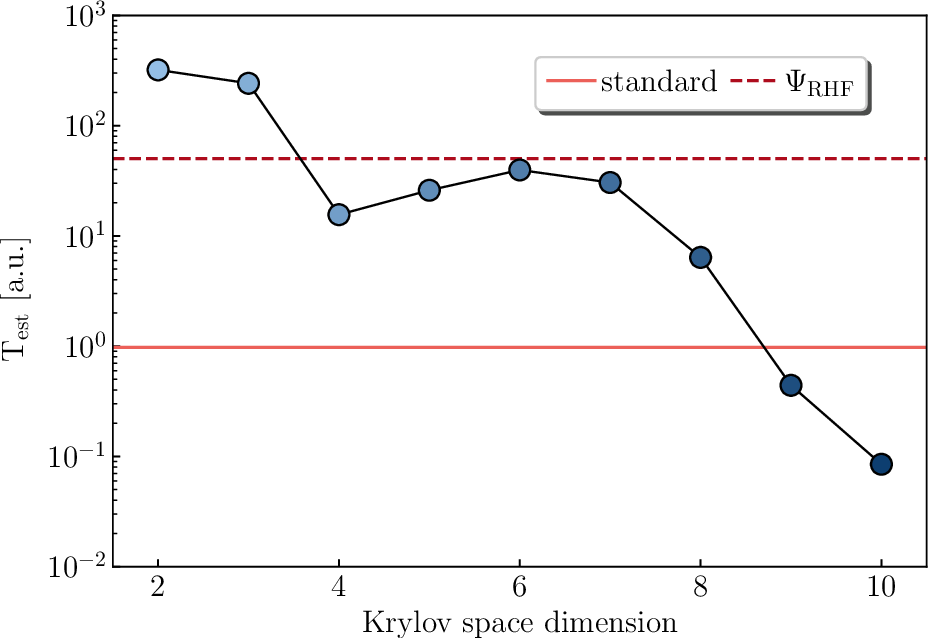}
\caption{
Estimate of the adiabatic time $\TIME{est}$ for singlet methylene at equilibrium geometry, using a (8e,6o) active space. $\TIME{est}$ is estimated for standard adiabatic state preparation (i.e. with RHF trial and the Fock operator as its parent, red solid line), for an RHF trial with the parent Hamiltonian construction proposed in this work (red dashed line), and for Krylov-space wavefunctions of increasing dimension (blue circles, with darker colors corresponding to higher $d$).
}
\label{fig:ch2_8e6o}
\end{figure}

In Figure~\ref{fig:ch2_8e6o}, we compute the adiabatic estimate $\TIME{est}$ for methylene, in an (8e,6o) active space of MOs with \aolabel{2s}, \aolabel{2p} character for carbon and \aolabel{1s} character for hydrogen, corresponding to 8 qubits. The trends seen in Figure~\ref{fig:h2o_4e4o} are seen also here: $\TIME{est}$ with RHF trial and parent Hamiltonian as proposed in this work is higher than for ``standard'' ASP and decreases with Krylov space dimension $d$, albeit non-monotonically. 

\revision{These results highlight the heuristic nature of our construction. Initial states which admit a natural parent Hamiltonian may observe that adiabatic paths resulting from our parent Hamiltonian construction are outperformed by paths constructed from a physically motivated parent Hamiltonian. This may suggest that out method is best suited for use with initial states that do not admit a natural parent Hamiltonian.}

\section{Discussion}
\label{sec:discussion}

In this Section, we discuss the computational cost, convergence properties, numerical stability, and effect of noise on the proposed algorithm.

\subsection{Computational cost}

In the present work, a quantum computer is used to measure a correlation matrix of the form Eq.~\eqref{eq:a_matrix}.
A classical computer is then used to identify a parent Hamiltonian, and a quantum computer is finally used to perform
an ASP calculation.
\begin{enumerate}
\item 
The first phase of the algorithm requires $O(m^2)$ measurements of Pauli operators, where $m$ is the number of Pauli operators in the qubit representation of the Hamiltonian. For the Born-Oppenheimer Hamiltonian, the focus of this work, $m = O(n^4)$ where $n$ is the number of qubits. The number of measurements can be reduced, e.g. by partitioning the operators $\pauli{i}$ into commuting families \cite{verteletskyi2020measurement,dutt2023practical}.
\item 
The second phase of the algorithm comprises (i) the diagonalization of the matrix $A$, at cost $O(m^3)$, to identify a kernel of dimension $\ell = \mbox{dim}(\kerspace(A)) \leq m$. Once an orthonormal basis $\{ \vett{e}_a \}_{a=1}^\ell$ for 
$\kerspace(A)$ is identified, the vector $\alphavec$ can be written as $\alpha_i = \sum_a (\vett{e}_a)_i x_a = (E\vett{x})_i$, where $\vett{x} \in \mathbb{R}^\ell$, and the cost function Eq.~\eqref{eq:cost_simplified} as
\begin{equation}
C[\vett{x}] = 4 (\tilde{\vett{b}} \cdot \vett{x})^2 \, \| \vett{x} \|^2  + 4 (\tilde{\vett{b}} \cdot \vett{x}) (\tilde{\betavec} \cdot \vett{x}) + \rho \sum_{ab} x_a x_b \tilde{\delta}_{ab}
\end{equation}
up to the constant $\| \betavec \|^2$. In the previous equation, $\tilde{\vett{b}} = E^T \vett{b}$, $\tilde{\betavec} = E^T \betavec$, and $\tilde{\delta} = E^T \bar \delta E$. 

These intermediate quantities do not depend on $\vett{x}$ and thus only need to be evaluated once. 
The computation of these intermediate quantities scales as $O(m \ell^2)$ and because they do not depend on $\vett{x}$ they only need to be calculated once. From that point onward, evaluating the cost function costs $O(\ell^2)$. Most iterative optimization methods would require at least this cost per iteration. Meanwhile, the heuristic initialization procedure which produces requires $\alphavec^*$ scales as $O(\ell)$

\item 
ASP for a time $\TIME{ASP}$ may be implemented, e.g., with a product formula involving $n_s$ steps,
\begin{equation}
\label{eq:trotter_1}
\hat{U}_{\mathrm{ASP}}(\TIME{ASP},n_s) = \prod_{k=0}^{n_s-1} \exp\left( - i \frac{T}{n_s} \hamiltonian{s_k} \right)
\;,\;
s_k = \frac{k}{n_s} 
\;.
\end{equation}
When the linear interpolation Eq.~\eqref{eq:linear_interpol} is used, 
$\hamiltonian{s_k} = (1-s_k) \parent[\alphavec] + s_k \hfinal$ is a linear combination of up to $m^2$ Pauli operators. This is because the parent Hamiltonian is obtained by shifting and squaring the proto-parent Hamiltonian, itself a linear combination of $m$ Pauli operators. 
The exponential in Eq.~\eqref{eq:trotter_1} can be implemented with e.g. a Suzuki product formula \cite{suzuki1990fractal} at the cost of $O(m^2 n)$ gates in depth $O(m^2 n)$. 
The computational cost of performing ASP with Eq.~\eqref{eq:trotter_1} is thus $O(m^2 n n_s)$ gates, 
with the number of Trotter steps, $n_s$, roughly scaling as $O(\TIME{ASP}^2/\eta)$, 
where $\eta$ is the target error on the output state of ASP \cite{childs2021theory}.
\end{enumerate}

\subsection{Convergence properties and numerical stability}

Several factors affect the ability to find an optimal $\alphavec$ by minimizing the cost function Eq.~\eqref{eq:cost}.
First, the cost function is not guaranteed to converge to $0$. Although the proto-parent $\protoparent[\alphavec]$ is a linear combination of the same Pauli operators in $\hfinal$, and approaches 
$\hfinal$ as $\alphavec \to \betavec$, the parent $\parent[\alphavec]$ is constructed with the folding operation in Eq.~\eqref{eq:parent}, i.e. by a shifting and squaring. Folding yields an operator that may be very different from $\hfinal$, e.g., because:
(i) $\parent[\alphavec]$ may contain Pauli operators that $\hfinal$ does not contain, or (ii) the spectrum of $\parent[\alphavec]$ may differ significantly from that of $\hfinal$ (due to the squaring, the former is non-negative, whereas the latter may have negative eigenvalues).

Furthermore, the cost function is approximated as in Eq.~\eqref{eq:cost_simplified} to avoid the expensive evaluation of a quartic function of $\alphavec$ and its challenging optimization (high-order polynomials lead to flat landscapes, that are in turn challenging to optimize). As a result, even if the optimization is successful, what is minimized is but an approximation to the exact cost function.

Due to these factors, it is difficult to guarantee that a sequence of trial wavefunctions converging to the ground state (e.g. Krylov states of increasing dimension $d$) will be associated with a sequence of parent Hamiltonians converging to the target Hamiltonian, nor that the corresponding $\TIME{est}$ will converge monotonically to zero and evolve differentiably with bondlength.
Encouragingly, the data in Figures~\ref{fig:h2o_4e4o},~\ref{fig:h2o_8e6o} and ~\ref{fig:ch2_8e6o} empirically illustrate that $\TIME{est}$ improves with Krylov space dimension $d$.

\subsection{Impact of shot noise}

In Section~\ref{sec:method}, we assumed that the correlation matrix $A$, Eq.~\eqref{eq:a_matrix}, is known exactly and has a null space, i.e., a set of eigenvectors with eigenvalue exactly equal to zero. On a classical simulator, the eigenvalues of the matrix $A$ are only known numerically within numeric precision, so the null space of $A$ is necessarily associated with eigenvalues below a user-defined threshold.
Furthermore, on a quantum computer, the elements of $A$ are measured by gathering a finite number of samples (or ``shots''), 
resulting in statistical fluctuations. Therefore, only a perturbed matrix $\tilde{A} = A + \Delta A$ is known, and the spectral norm $\| \Delta A \|$ of the perturbation decreases with the number of shots. \footnote{The spectral norm is bounded by the Frobenius norm 
$\| \Delta A \|_\mathrm{F}^2 \leq \mbox{Tr}[\Delta A^\dagger \Delta A]$. Furthermore, $\mbox{Tr}[\Delta A^\dagger \Delta A] = \sum_{ij} \Delta a_{ij}^2$ may be crudely approximated with $\sigma^2 m^2/N_{\mathrm{shot}}$ assuming that matrix elements $\Delta a_{ij}$ are estimated with $N_{\mathrm{shot}}$ shots each, and are} independently and identically distributed with variance $\sigma^2/N_{\mathrm{shot}}$.
We remark that the bound in this footnote is rather crude, and that a tighter bound can be obtained following e.g. Theorem 4 of \cite{lee2023sampling}.

The null space of $\tilde{A}$ can be identified by computing its eigenvalues, $\{ \mu_k \}_{k=1}^m$, and retaining those below a threshold $\delta$, $\mu_1 \leq \dots \leq \mu_{\ell} \leq \delta$. Let also $\gamma= \mu_{\ell+1}$ be the smallest eigenvalue above $\delta$. 

A natural question is whether the null space of $\tilde{A}$ is close to the null space of $A$. An answer to this question comes from a special form of the Davis-Kahan theorem \cite{davis1970rotation,rajendra1997matrix,kirby2024analysis}: given two Hermitian matrices $A$ and $\tilde{A}$, and their spectral projectors
$\Pi_K$ and $\tilde{\Pi}_{\tilde{K}}$ onto subsets of the real axis with $\min(K)-\max(\tilde K) \geq \eta$ for some gap $\eta$, then
$\| \Pi_K \tilde{\Pi}_{\tilde{K}} \| \leq \| A- \tilde{A} \| / \eta$. Applying the Davis-Kahan theorem to the correlation matrices $A$ and $\tilde{A}$ of the present Section, and to the sets $\tilde{K} = (-\infty,\delta]$ and ${K} = [\gamma,+\infty)$, we obtain
\begin{equation}
\| \Pi_K \tilde{\Pi}_{\tilde{K}} \| \leq \frac{ \| \Delta A \| }{ \gamma-\delta }
\;.
\end{equation}
The projector on the subspace of ${A}$ with eigenvalues above $\gamma$ is approximately orthogonal to the projector on the subspace of $\tilde{A}$ with eigenvalues below $\delta$ (i.e. its null space), provided that $\| \Delta A \| \ll \gamma-\delta$. As the perturbation decreases in spectral norm with increasing number of shots, the algorithm proposed here can be regarded as robust under shot noise provided the covariance matrix $A$ is gapped.

\section{Conclusion and outlooks}
\label{sec:conclusion}

In this work, we presented a method to compute a parent Hamiltonian for a generic trial wavefunction $\psitrial$ that can be prepared by a quantum circuit. We propose to use a quantum processor to estimate a correlation matrix $A$, whose null space specifies a vector space of proto-parent Hamiltonians for whom $\psitrial$ is an eigenstate, though not necessarily the ground state. A parent Hamiltonian is then produced by shifting and squaring the proto-parent Hamiltonian, which can then be used for adiabatic state preparation (ASP).

Numerical experiments on water and singlet methylene have identified encouraging behavior along with technical challenges: the $\TIME{est}$ obtained for Krylov wavefunctions decreases with Krylov space dimension, whereas the numerical optimization of $\alphavec$ for the RHF wavefunction yields $\TIME{est}$ above that of standard ASP (i.e. with RHF trial wavefunction and the Fock operator as the parent Hamiltonian). One could interpret these results to mean that the methods put forth in this work are most promising when applied to trial wavefunctions that do not already admit a natural parent Hamiltonian.

The method detailed in this work factorizes into three components: the choice of the Pauli operator support $\pauliset$, the conversion of a proto-parent Hamiltonian into a parent Hamiltonian via spectrum folding, and the optimization of a cost function based on the distance between parent and target Hamiltonian.

In general, if $|\pauliset|$ scales polynomially with the number of qubits $n$, one cannot guarantee that $A$ has a non-trivial null space, i.e., $\mbox{dim}(\kerspace(A)) \geq 1$.  If one imposes additional conditions onto $\psitrial$, more rigorous claims can be made about the performance of our approach. For example, previous work considered the construction of a proto-parent Hamiltonian assuming that $\psitrial$ is an eigenstate of an unknown local Hamiltonian \cite{qi2019determining}: under such a hypothesis, a polynomially large set $\pauliset$ can be chosen such that $\protoparent[\alphavec]$ is uniquely defined and exactly recoverable. Other previous works consider similar situations, where the trial state is assumed to obey specific formulations of the eigenstate thermalization hypothesis \cite{moudgalya2020large, eth_eigenstate}. Because we do not impose such assumptions on the trial state of our method, we do not recover any of these previously known performance guarantees, and the application of our method is subject to the user's ability to provide a suitable set $\pauliset$ for the specific situation at hand. On the other hand, the success of our procedure can be easily verified (success if $\mbox{dim}(\kerspace(A)) \geq 1$) and 
our procedure is robust under shot noise if the covariance matrix is gapped.

When successful, our method produces a subspace of parent Hamiltonians parameterized by a vector $\alphavec$. The \revision{shifting and squaring} operation guarantees 
that $\psitrial$ is an eigenstate of $\parent[\alphavec]$ with eigenvalue $0$, and that 
$\parent[\alphavec] \geq 0$. 
However, we cannot exclude that the eigenvalue $0$ is degenerate or that the spectral gap of $\parent[\alphavec]$ vanishes in the thermodynamic limit of large system size, in which cases the adiabatic theorem would not hold.
The latter situation occurs if the proto-parent Hamiltonian has a continuum of eigenstates (typically away from both the ground and highest excited state) and $\psitrial$ lies in such a continuum.  
Encouragingly, in this work, we observed that $\parent[\alphavec]$ is gapped and its ground state unique.

In this work we determined the parameter $\alphavec$ by minimizing the Frobenius distance between $\parent[\alphavec]$ and $\hfinal$, used as the cost function $C[\alphavec]$ of a classical numerical optimization problem. This choice is appealing because the Frobenius distance is efficiently computable on a classical device. 
Furthermore, if $C[\alphavec] \simeq 0$, the difference $\op{V} = \parent[\alphavec] -\hfinal$ is a small operator (in the sense of the Frobenius norm) and the time-dependent Hamiltonian $\aspham{s} = \hfinal + (1-s) \op{V}$ can be written as the target Hamiltonian plus a small perturbation: in this regime, if the target Hamiltonian is gapped, so is the time-dependent Hamiltonian (by a simple application of Rayleigh-Schr\"{o}dinger perturbation theory).
However, (high-energy) excited states contribute significantly to the Frobenius distance, through the trace operation; moreover, $C[\alphavec]$ may not decrease to $0$ upon optimization when folding is used.
\revision{Important areas for further work include exploring alternative cost functions for parent Hamiltonian optimization, optimizing the choice of operators to include in $\pauliset$, studying how our method can be optimized for highly correlated or small gapped systems, and investigating the performance of ASP using constructed parent Hamiltonians with optimized scheduling functions.}

\revision{We acknowledge that the squaring of $\protoparent$ to produce $\parent$ poses a limitation for our method, as this leads to a number of terms in $\parent$ which scales quadratically in $|\pauliset|$ and a doubling of the maximum Pauli weight with respect to the elements of $\pauliset$. The increase in the Pauli weight and number of terms in $\aspham{s}$ will result in each step of time evolution being more costly to implement. Thus, for our method to be practically advantageous the initial state and constructed $\parent$ must require proportionally fewer time steps such that the total circuit depth remains less than what is needed to perform ASP from a physically motivated parent Hamiltonian.}

We believe the present study will be a useful contribution to the understanding and development of adiabatic quantum computing for electronic structure and other applications in many-body quantum mechanics.

\section*{Acknowledgments}

We thank W. Kirby, M. Tran, and J. Garrison for helpful feedback on the manuscript.

\appendix

\section{Details of the parent Hamiltonian construction}

\subsection{Cost function of the optimization}
\label{app:cost_function}

To simplify the optimization of the cost function Eq.~\eqref{eq:cost}, we first observe that
\begin{equation}
\sum_{ij} \alpha_i \alpha_j \pauli{i} \pauli{j}
= \left[ \sum_i \alpha_i^2 \right] \op{I} + \sum_{i<j} \alpha_i \alpha_j \{ \pauli{i} , \pauli{j} \}
\;.
\end{equation}
If the set $\pauliset$ consisted of anticommuting Pauli operators, the second term would vanish: motivated by this observation, we (i) approximate the parent Hamiltonian (exclusively to optimize
$\alphavec$) as
\begin{equation}
\parent[\alphavec] 
\simeq \left[ \alphavec \cdot \alphavec 
+ (\alphavec \cdot \vett{b})^2 \right] \op{I}
- 2 (\alphavec \cdot \vett{b}) \sum_i \alpha_i \pauli{i}
\;,
\end{equation}
and (ii) add a penalty term to the cost function, of the form $\rho \sum_{ij} \alpha_i \alpha_j \delta_{ij}$ with $\rho \geq 0$ and
\begin{equation}
\delta_{ij}
=
\left\{
\begin{array}{ll}
1 & \mbox{if $i \neq j$ and $\{ \pauli{i} , \pauli{j} \} \neq 0$} \\
0 & \mbox{otherwise} \\
\end{array}
\right.
\;.
\end{equation}
Furthermore, we can write the target Hamiltonian without approximations as 
\begin{equation}
\hfinal = \sum_i \beta_i \pauli{i} 
\;,
\end{equation}
leading to
\begin{equation}
\| \parent[\alphavec] - \hfinal \|_\mathrm{F}^2
\simeq
\left( \alphavec \cdot \alphavec + ( \alphavec\cdot \vett{b})^2 \right)^2
+
\| 2 ( \alphavec\cdot \vett{b}) \alphavec + \betavec \|^2
\;.
\end{equation}
The term $\alphavec \cdot \alphavec + ( \alphavec\cdot \vett{b})^2$ only contributes a shift to the spectrum of $\parent[\alphavec]$, so we do not include it here. This results in the approximate cost function in Eq.~\eqref{eq:cost_simplified}.

\subsection{Initial point of the optimization}
\label{app:cost_initial}

Minimizing the cost function Eq.~\eqref{eq:cost_simplified} is a non-convex optimization problem, that may converge to a local minimum. In this situation, it is important to start the optimization from a reliable and efficiently computable initial point. To this end, let us introduce an orthonormal basis 
$\{ \vett{e}_i \}_{i=1}^\ell$ for $\kerspace(A)$, such that
\begin{equation}
\vett{b} = b_1 \vett{e}_1 + \vett{b}_\perp
\;,\;
\betavec = \sum_{i=1}^\ell \beta_i \vett{e}_i + \betavec_\perp
\;,\;
\alphavec = \sum_{i=1}^\ell x_i \vett{e}_i
\end{equation}
i.e., the first element of the basis is the projection of $\vett{b}$ onto $\kerspace(A)$, and
$\vett{b}_\perp/\betavec_\perp$ are the projections of $\vett{b}/\betavec$ outside $\kerspace(A)$.
We also assume that $b_1 \neq 0$. The function $\| 2 ( \alphavec\cdot \vett{b}) \alphavec + \betavec \|^2$ then takes the form
\begin{equation}
\| 2 ( \alphavec\cdot \vett{b}) \alphavec + \betavec \|^2
= \sum_{i=1}^\ell \left( 2 x_1 b_1 x_i + \beta_i \right)^2
\;.
\end{equation}
Note that, if $b_1 = 0$, this quantity is independent of $\alphavec$ and equal to $\| \betavec \|^2$.
Otherwise, it takes the minimum value $(2x_1^2 b_1 + \beta_1)^2$ at
\begin{equation}
\alphavec^* = x_1 \vett{e_1} - \frac{2}{x_1 b_1} \sum_{i=2}^\ell \beta_i \vett{e}_i
\;.
\end{equation}
If $\beta_1/b_1<0$, the choice $x_1 = \sqrt{- \beta_1/b_1}$ leads to $\| 2 ( \alphavec\cdot \vett{b}) \alphavec + \betavec \|^2 = 0$. Otherwise, one can treat $x_1$ as a free parameter and minimize $C[\alphavec^*]$ versus $x_1$.

\subsection{Treatment of Hamiltonian symmetries}

As discussed in the main text, $\mathbb{Z}_2$ symmetries of the target Hamiltonian (e.g. $(-1)^{\hat{N}_\sigma}$ with $\sigma = \alpha,\beta$ and molecular point-group symmetries isomorphic to $\mathbb{Z}_2^{\times n}$ such as $\mathrm{C_{2v}}$) can be exactly enforced via the qubit tapering. Other symmetries exist, which are not of $\mathbb{Z}_2$ type (e.g. ${\hat{N}_\sigma}$ with $\sigma = \alpha,\beta$, total spin $\hat{S}^2$ and $\hat{S}_z$, general molecular point-group symmetries such as $\mathrm{C_{\infty v}}$ or $\mathrm{D_{\infty h}}$). It is desirable to produce a parent Hamiltonian that commutes with such symmetries, which can be accomplished as follows.

First, let us recall that an operator $\hat{X} = 0$ is equal to zero if and only if its Frobenius norm vanishes, $\| \hat{X} \|_\mathrm{F}^2 = \mbox{Tr}[\hat{X}^\dagger \hat{X}] = 0$. Consider now 
\begin{equation}
\hat{X} = 
\left[ \hat{B} , \protoparent[\alphavec] \right] = 
\sum_i \alpha_i [\hat{B} , \pauli{i} ] \equiv \sum_i \alpha_i \, \hat{C}_i
\end{equation}
where $\hat{B}$ is any symmetry of $\hfinal$, i.e., any Hermitian operator that commutes with $\hfinal$. The condition $\| \hat{X} \|_\mathrm{F}^2 = 0$ takes the form
\begin{equation}
0 = 
\sum_{ij} \alpha_i \alpha_j \mbox{Tr}[ \hat{C}^\dagger_i \hat{C}_j ] = 
\sum_{ij} \alpha_i \alpha_j \mbox{Tr}\left[ \frac{ \hat{C}^\dagger_i \hat{C}_j + \hat{C}^\dagger_j \hat{C}_i }{2} \right] = 
\sum_{ij} \alpha_i \alpha_j B_{ij}
\;,\;
\end{equation}
i.e. $\alphavec \in \kerspace(B)$. In the presence of $k$ Hamiltonian symmetries $\{ \hat{B}_s \}_{s=0}^{k-1}$, the requirement for $\protoparent[\alphavec]$ to commute with all $k$ symmetries is that $\alphavec$ lies in the intersection between the null spaces of the corresponding matrices $B^{(s)}$,
\begin{equation}
\alphavec \in \bigcap_{s=0}^{k-1} \kerspace\left( B^{(s)} \right)
\;.
\end{equation}

\subsection{Hamiltonian prefactor optimization}

In Section \ref{alpha_optimization} we illustrated how $\alphavec$ can be optimized to minimize 
$\| \parent[\alphavec] - \hfinal \|_\mathrm{F}^2$. However, once an $\alphavec$ has been chosen, one has an additional degree of freedom to choose a scaling factor $c \geq 0$ for the parent Hamiltonian and to then optimize Eq.~\eqref{eq:t_est} as a function of that parameter only. 
Focusing for simplicity on the linear path Eq.~\eqref{eq:linear_interpol}, for which
\begin{equation}
\label{eq:linear_derivative}
\frac{d \hat{H}}{ds}(s) = 
\aspham{s} = \hfinal - c \parent
\;,
\end{equation}
the estimate Eq.~\eqref{eq:t_est} takes the form
\begin{equation}
\TIME{est} = \max_s f_s(c) \geq \max \{ f_0(c),f_1(c) \}
\;,\;
f_s(c) = \max_j 
\frac{
\left| \langle \phi_j(s) | \hfinal - c \parent | \phi_0(s) \rangle \right|}
{ \left( \lambda_j(s) - \lambda_0(s) \right)^2 }
\;.
\end{equation}
At $s=0$, where $\hamiltonian{0}= c \parent$, one has
\begin{equation}
f_0(c) = \max_j \frac{ \left| \langle \phi_j(0) | \hfinal | \phi_0(0) \rangle \right|}
{ \left( \lambda_j(0) - \lambda_0(0) \right)^2 } = \frac{a_0}{c^2}
\;.
\end{equation}
Note that the $c^2$ at the denominator arises because the eigenvalues $\lambda_j(0)$ of $\hamiltonian{0}= c \parent$ are scaled by a factor of $c$.
At $s=1$, where $\hamiltonian{1} = \hfinal$, on the other hand,
\begin{equation}
f_1(c) = \max_j \frac{ \left| \langle \phi_j(1) | c \parent | \phi_0(1) \rangle \right|}
{ \left( \lambda_j(1) - \lambda_0(1) \right)^2 } = a_1 c
\;.
\end{equation}
Therefore, $\TIME{est} \geq \max \{ a_0/c^2 , a_1 c \}$ is lower-bounded by $O(1/c^2)$ for $c \to 0$ and
$O(c)$ for $c\to \infty$. An estimate for the optimal value of $c$ is the intersection point of the curves $f_0(c)$ and $f_1(c)$, i.e., $c^* = \sqrt[^3]{a_0/a_1}$.

\section{Details of quantum chemistry calculations}
\label{eq:app_chem}

Our calculation for H$_2$O and CH$_2$ used a minimal STO-6G basis set \cite{hehre1969self}, and isosceles triangular geometries.
For H$_2$O, we employed an angle of $\theta = 104.4776$ degrees \cite{hoy1979precise} and we rescale the equilibrium bond lengths with a parameter $R$ varying between 0.6 and 2.3.
For methylene, we employed the \spectro{1}{\mathrm{A}_1} equilibrium geometry \cite{petek1989analysis} with an angle of $\theta = 102.373$ degrees
and a bondlength of $R = 1.107 \mathrm{\AA}$.
We carried out RHF calculations using PySCF, enforcing $\mathrm{C_{2v}}$ symmetry. 
We then extracted an active-space Hamiltonian of the form
\begin{equation}
\hat{H} = E_0 + \sum_{pr,\sigma} h_{pr} \crt{p\sigma} \dst{r\sigma} + \sum_{prqs,\sigma\tau} \frac{(pr|qs)}{2}
\crt{p\sigma} \crt{q\tau} \dst{s\tau} \dst{r\sigma}
\;,
\end{equation}
using PySCF's CASCI library to extract the coefficients $E_0$, $h_{pr}$, and $(pr|qs)$ for the selected active-space MOs.

We carried out active-space MP2 calculations with PySCF, and Krylov space calculations with an in-house code based on
IBM's Qiskit library, that formed the basis vectors $\hat{H}^p | \psi_{\mathrm{RHF}} \rangle$ by multiplying the operator
$\hat{H}$ times the vector $| \psi_{\mathrm{RHF}} \rangle$.

\bibliographystyle{tfo}
\bibliography{main}
\end{document}